# How IFRS Affects Value Relevance and Key Financial Indicators?
# Evidence from the UK

## Yhlas SOVBETOV[a]

a.  *Department of Economics, Istanbul University, Istanbul, Turkey*

**Abstract:** This paper has two contributions to the International Financial Reporting Stands (IFRS) adoption literature. First is the scrutinizing impact of IFRS adoption on value relevance in the UK with TEST-A analysis under the $H_{01}$ hypothesis. The second contribution is capturing the impact of IFRS adoption on key financial indicators of firms with the TEST-B analysis that hypothesizes $H_{02}$.The statistical differences of items of two different reporting standards are examined with non-parametric tests as all input variables failed the Shapiro-Wilk and Lilliefors normality tests in TEST-A. The finding rejects the $H_{01}$ hypothesis for BvMv, and agrees that IFRS has impact on value relevance. Besides, Ohlson's (1995) model documents that the coefficient of dummy variable (MODE) is positive. Therefore, the analysis concludes that IFRS has positive impact on value relevance. The aftermath of TEST-B rejects the $H_{02}$ hypothesis for all profitability ratios (ROE, ROCE, ROA, PM) and gearing ratios (GR). It concludes that profitability and gearing ratios are affected by IFRS adoption, whereas efficiency-liquidity ratios are not. Also, in Forward Stepwise regression analysis only ROCE, ROA, and PM ratios show significant results. The analysis documents positive and significant impact of IFRS on these three ratios.

## 1. Introduction

The rationale of financial reporting is to supply transparent financial information and outlook about a firm to investors and the public. As all firms in the world do not stand on the same base of accounting and reporting framework, a healthy comparability by investors or interested parties was nearly impossible. Besides in recent years international financial transactions have been boosted up with the increase of competitiveness of companies over the world, and as Lee et al (2010) states, the stock prices of firms became more sensitive to these international activities. This globalization encouraged local investors to seek investment opportunities in outside of country boundaries. But interpretation and understanding international financial transactions were a big concern for these investors due to dissimilarities of accounting standards on reporting at international level. Therefore, since 1970, the International Accounting Standards Board (IASB) has been working with the cooperation of European Union (EU), on harmonization of different countries' accounting models, and creating a unique international framework to mitigate these dissimilarities and concerns. Also, this international framework would help to increase simplicity and flexibility in understanding and reporting the financial information. In a word, as Silva, Do Couto, and Cordeiro (2009) stated, the necessity of global accounting framework is emerged by progress and expansion of business transactions at international level.

IFRS is a set of accounting principles, which was firstly introduced in 2001 by IASB, and still today, it has been accustomed by more than hundred countries around the world in preparation of their public firms' financial reports and statements (Tanko, 2012). According to Terzi, Oktem, and Sen (2013), all 27 members of the EU, as well as many African, Asian, and American countries are integrated into the IFRS for their local GAAPs, and have started to base their report publications on a new regime.

The adoption and implementation process of a newly designed accounting framework International Financial Reporting Standards (IFRS) in the EU was a major financial event as the first regulatory basis in the world's finance history ever (Hung and Subramanyam, 2004). Especially, the aftermath of the first successful experience of the IFRS by some EU countries in the period of 1998-2002, led the new framework to become mandatory by law in June 2002, for public listed companies in the EU. According to that regulation, the public listed companies (PLC) are required to construct and design their balance sheets and reports under this new framework IFRS (Callao, Jarne, Lainez, 2007). This adoption of the EU, led to the framework to be accepted and accustomed at worldwide level by other countries such as US, Canada, Australia, and many Asian and African countries as well.

Since the past decade, many studies have investigated IFRS adoption and its impact on financial indicators in different countries such as Hung and Subramanyam (2004) in Germany; Jermakowicz (2004) and Haverals (2005) in Belgium; Callao, Jarne, and Lainez (2007) in Spain; Agca and Aktas (2007), as well as Terzi, Oktem, and Sen (2013) in Turkey;

Rahmanova (2009), and Harris, et al. (2013) in the US; Lantto and Sahlstrom (2009) in Finland; Silva, Do Couto, and Cordeiro (2009) in Portugal; Brochet, Jagolinzer, and Riedl (2011), as well as Punda (2011) in the UK; Klimczak (2011) in Poland; Prochazka (2010), beside Palka and Svitakova (2011) in Czech Republic; Outa (2011) in Kenya; Tanko (2012) in Nigeria; Csebfalvi (2012) in Hungary; Hilliard (2013) in Canada; Tsalavoutas, Andre, and Evans (2012) in Greece; Adzis (2012) in New Zealand; Jermakowicz and Gornik-Tomaszewski (2006), Gebhardt and Novotny-Farkas (2010), Lee, Walker, and Christensen (2008), Blanchette, Racicot, and Girard (2011), as well as Kubickova and Jindrichovska (2012) in the EU. The majority of these studies have documented a significant impact of IFRS on balance sheet figures and key financial ratios, but few studies have concluded no evidence of discernible changes. A brief breakdown of empirical studies is presented in Table 1.

## 2. Literature Review

Hung and Subramanyam's (2004) study examines the effects of adopting IFRS on financial statements and value relevance for 84 German firms during 1998-2002. They compare accounting numbers reported under German accounting standard (HGB) and under IFRS for the same firm-years, and find that the new regime has greater emphasis on the balance sheet and fair valuation, but less focus on income smoothing. Besides they use Ohlson's (1995) model in value relevance analysis. The model regresses the price (market value of equity) with book value of equity and net income. Under this model, they document that IFRS improves the value relevance of book value and total assets, but fails in valuing net income. They also underline that HGB is conservative and income-smoothing oriented, while IFRS is fair-value and balance-sheet oriented.

On the other hand, Callao et al (2007) investigates the adoption of IFRS in Spain. They sampled 35 listed firms for 2004-2005 periods, and run normality tests (Kolmogorov–Smirnov, Shapiro–Wilks) before t-tests, and non-parametric (Wilcoxon signed-ranks) tests for non-normal variables. They find that local comparability has worsened with the IFRS, and no improvements in the relevance of financial reporting, because they observe the gap between book and market values are widened under IFRS. They explain this as accounting policies may show balance conservatism when the market value exceeds the book value, and common-law countries (like US and UK) are less conservative than code-law countries (like Spain, Italy, Russia). The IFRS is close to common-law that shows less conservatism than the SAS because it is code-law based.

In the case of Turkey, Terzi et al (2013) examine impact of adopting IFRS of 140 manufacturing firms listed in Istanbul Stock Exchange during 2004-2006 years. They check the normality of variable with Kolmogorov-Smirnov and Shapiro-Wilks tests, and apply Wilcoxon signed-ranks test to identify the relationship between the non-parametric variables. They employ logistic regression model for empirical analysis, and find that IFRS has significant impact on inventories, fixed assets, long-term liabilities, and shareholder's equity accounts. Besides, they document that current ratios, asset turnover ratios, and financial leverage ratios are significantly affected by new regime. However, they find no evidence of improvements in value relevance. Likewise, Agca and Aktas (2007) analyze adoption of Turkish firms during 2004-2005 with t-test. They find that only current ratio and net asset turnover ratios are affected significantly with IFRS.

Moreover Lantto and Sahlstrom (2009) study IFRS adoption of 91 firms listed in Helsinki Stock Exchange during 2004-2005. They mainly focus on three different key economic dimensions of a firm, i.e. profitability, financial leverage and liquidity. They run univariate analysis and Wilcoxon test, and observe that profitability and financial leverage ratios (OPM, ROE, ROCE, GR) are increased, while liquidity ratios (QR and CR) are decreased under IFRS.

Also, Silva et al (2009) measure the impact of the application of IFRS in Portugal with 39 firms listed in Lisbon Stock Exchange. Firstly, they use descriptive statistics to examine the means of selected ratios (Gearing Ratio, PER, and EPS). Afterwards, they apply K-Means Cluster Analysis which groups similar firms and adopts the Euclidian distance to measure the distance, or difference between each firms. They find that the balance sheet variables such as intangible assets, fixed tangible assets, investments, equity and liabilities are significantly affected. However, income statement variables appear low significant. They also use linear regression model computing the t, F and $R^2$ statistics, and could not document a clear evidence of accounting variations.

As an example for common-law adoptions, Punda (2011) samples 101 British firms listed in London Stock Exchange during 2005. Punda tries to confirm Lantto and Sahtstrom's (2009) statement that there are no substantial differences between IFRS and accounting standards under common law regime. To do this, he selects five financial ratios (OPM, ROE, ROIC, CR, PE), and calculates the difference by subtracting a median value of every financial ratio under UK GAAP from the median values of financial ratio under IFRS.

Further, he tests the statistical significance of the differences by utilizing non-parametric Wilcoxon Signed-Rank Test. As a result, he finds that despite widely believed similarities between UK GAAP and IFRS, there are significant differences in accounting numbers. He documents that all profitability ratios are increased after transition to new regime. Similarly, liquidity ratios increased substantially, but less significant. On the other hand, PE ratio (market based Price to Earnings ratio) is slightly decreased after transition. He concludes that obtained results of increase in profitability ratios and decrease in PE ratio indicate very high income statement profits under the new regime. Likewise, Brochet et al (2011) investigate the adoption effects in same country. Despite to Punda (2011), they mainly focus on comparability improvements of financial statements and changes in information quality by multivariate insider trading analysis. They observe that abnormal returns of both insider purchase and analyst recommendation upgrades are reduced with IFRS transition. So, they conclude that IFRS reduces private information by enhancing the comparability of financial statements.

In the case of Canada, Blanchette et al (2011) try to capture the Canadian GAAP-IFRS transition impact on 9 firms during 2008-2009 with univariate analysis and linear regression model. The results of descriptive statistics exhibit non-normal distribution of means and medians. Therefore, they conduct non-parametrical tests. As a result, they find no significant difference between medians of all ratios (Profitability, Liquidity, Leverage, Coverage), but an increase in volatility of all of these ratios under IFRS.

Another researcher for same country is McConnel1 (2012) who samples 50 Canadian mining firms, and examines IFRS adoption impact over 2010-2011. He focuses on same ratios as Blanchette et al (2011), and utilized Ryan-Joiner, Levene's and Wilcoxon signed rank tests, as variables are not normally distributed. As a result, similar to Blanchette et al (2011), he finds that no statistically significant differences existed in the dispersion of the ratios. However, he observes significant differences in the central tendency of three of the ratios: quick ratio, return on assets, and comprehensive return on assets. But these results cannot be generalized to all Canadian firms, as the study only examines mining industry. On the other hand, Hilliard (2013) investigates the value relevance improvements for same country. She samples 39 Canadian firms during 2009-2010, and applies univariate, bivariate (correlation & collinearity), and multivariate (multiple regression) analyses. But she finds no evidence of improvements in value relevance. Similarly to Blanchette et al (2011), she documents that volatility of ratios are increased under new regime.

## Table 1: The Breakdown of Empirical Studies

| Reference | Origin | Sample | Period | Model | Result |
|---|---|---|---|---|---|
| **Callao, Jarne, Lainez(2007)** | Spain | 35 Firms | 2004-2005 | Kolmogorov–Smirnov Shapiro–Wilks test Wilcoxon signed-ranks | **Increases:** CR,LTD,TD,ROE,GR **Decrease:** Debtors, Equity, OI, Solvency, ROA, QR |
| **Jermakowicz (2004)** | Belgium | 20 Firms | 2003 | Comparative Analysis | **Increase:** Earnings Volatility, Transparency, Comparability |
| **Palka & Svitakova (2011)** | Czech Republic | 115 Firms | 2004-2005 | Comparative Analysis | **Increase:** GR **Decrease:** ROE |
| **Kubickova & Jindrichovska (2012)** | Czech Republic | 18 Firms | 2005 | t-Test | No Significant Overall Changes |
| **Lantto & Sahlstrom (2009)** | Finland | 91 Firms | 2004-2005 | Descriptive Statistics Wilcoxon signed-rank | **Increase:** OPM,ROE,ROCE,GR **Decrease:** QR, CR, PE |
| **Arouri, Levy, Nguyen (2010)** | France | 40 Firms | 2004 | Longitudinal Analysis | **Increase:** ROE, NI, GR **Decrease:** ROA **No Change:** Transparency |
| **Hung & Subramanyam (2004)** | Germany | 84 Firms | 1998-2002 | Multiple Regression | **Increase:** TA, BV relevance **Decrease:** NI relevance. |
| **Tsalavoutas, Andre & Evans (2012)** | Greece | 159 Firms | 2001-2008 | Multiple Regression F-Test | **Increase:** Equity, NI **Decrease:** GR, Liquidity |
| **Csebfalvi (2012)** | Hungary | 260 Firms | 2006-2008 | Logistic Regression | **Increase:** Dividend, CR, GR **Decrease:** ROCE, Profitability **No Change:** EPS |
| **Beke (2011)** | Hungary | 325 Firms | 2006-2007 | Logistic Regression | **Increase:** GR, NP Volatility, Transparency **Decrease:** OCF, Solvency, Profitability |
| **Klimczak (2011)** | Poland | 159 Firms | 2000-2008 | Descriptive Statistics Multiple Regression Z-Test | **Increase:** Earnings Volatility **No Change:** Very less impacts on balance sheet items |
| **Silva, Do Couto, & Cordeiro (2009)** | Portugal | 39 Firms | 2004-2005 | Descriptive Analysis Multivariate Statistics K-Means Cluster Linear Regression | **Increase:** TL, PAT, OP **Decrease:** Eq, OC, GR, PE |
| **Terzi, Oktem, & Sen (2013)** | Turkey | 140 Firms | 2003-2005 | Kolmogorov-Smirnov Shapiro-Wilks tests Wilcoxon signed-ranks Mann-Whitney U test Logistic Regression | **Increase:** CR, LQR, ROA, ROE, GR **No Change:** ST, CA, CL, BV/MV |
| **Agca & Aktas (2007)** | Turkey | 147 Firms | 2004-2005 | t-Test | CR and NAT affected significantly |
| **Punda (2011)** | UK | 101 Firms | 2005 | Wilcoxon Signed-Rank Test | **Increase:** CR, ROE, PM, ROCE, CA, Operating Income, Net Income **Decrease:** PE, Revenue, CL, Equity |
| **Brochet, Jagolinzer, & Riedl (2011)** | UK | 663 Firms | 2003–2006 | Descriptive Statistics Multiple Regression | **Increase:** Earnings and Stock Volatility, Size of Purchase, **Decrease:** Size of Firm |
| **Gebhardt & Novotny-Farkas (2010)** | 12 EU Countries | 90 Banks | 2000-2007 | Descriptive Statistics Multiple Regression F-Test | **Increase:** EBIT **Decrease:** Loan Loss Provisions **No Change:** NPL, RegCap |
| **Jermakowicz & Gornik-Tomaszewski (2006)** | EU | 410 Firms | 2004 | Survey, T-test Spearman's Rank-Order Kendall's Tau-b | **Increase:** Equity, Extra Costs **Decrease:** Cost of Equity **No Change:** EPS, Total Assets, Revenue, Number of Employees |
| **Blanchette, Racicot, Girard (2011)** | Canada | 9 Firms | 2008-2009 | Descriptive statistics Linear Regression | **Increase:** Volatility of Ratios **No Change:** Means & Medians of Ratios |
| **McConnell (2012)** | Canada | 50 Firms | 2010-2011 | Wilcoxon Signed Rank Levene's Test Normality Tests Ryan-Joiner test | **Decrease:** CR, QR, ROA **No Change:** DebtR, EquityR, NAT |
| **Hilliard (2013)** | Canada | 39 Firms | 2009–2010 | Descriptive Statistics Bivariate Analysis Multivariate Analysis Shapiro-Wilk test Pearson Correlation Collinearity Statistics Multiple Regression | **Increase:** NI, GR **Decrease:** Earnings Volatility **No Change:** Value Relevancy |
| **Rahmonova (2009)** | USA | 25 Firms | 2005-2007 | Descriptive Statistics Comparative Analysis | **Increase:** ROA, ROE, PM, NAT, **Decrease:** CR, WC, **No Change:** GR |
| **Outa (2011)** | Kenya | 32 Firms | 1995-2004 | Descriptive Statistics Multiple Linear Regression | **Increase:** Earnings Volatility, Profitability **Decrease:** Value Relevancy |
| **Tanko (2012)** | Nigeria | 220 Firms | 2007–2010 | Descriptive Statistics Multiple Linear Regression T-test | **Increase** Profitability **Decrease:** CF, NI, Liquidity , Earnings Volatility |
| **Adzis (2012)** | Asia Pacific Region | 62 Banks | 1995-2009 | Descriptivr Statistics Correlation Matrix Robustness Test T-test Wilcoxon signed rank | **Increase:** Volatile of Earnings **Decrease:** Income Smoothing **No Change:** pro-cyclical behaviour of loan impairments |

The remainder of the paper is structured as follows. In section three, the methodology of this study is covered. In section four, the findings of assigned analyses are discussed and interpreted. Finally, the conclusion is structured in section five, followed by references and appendices.

# 3. Data and Methodology

## 3.1. Sample Selection

The 80 largest firms are selected from FTSE 100 index to scrutinize impact of IFRS on value relevance in the TEST-A analysis. However, the sample size has decreased to 65 firms in the TEST-B analysis that aims to gauge impact of IFRS on key financial indicators. In data selection, mainly the Fame, Morningstar and Company Intelligence databases are utilized. Firstly, official balance sheets and income statements of 2003-2006 periods for sampled firms are derived from mentioned databases. Subsequently ten key financial ratios and five balance sheet items are selected as input data variables for this study.

On the other hand, characteristics of selected firms differ due to their market structure. But it is not a limitation for pursuing this research. On the contrary, despite of examining a specific market, analyzing only the largest firms from different markets is more logic, and the IFRS impact is anticipated to be more obvious as those firms are important actors on the markets. Therefore, the industrial base of the sample firms is structured by different sectors such as financial, consumption goods, manufacturing, media, extractive, and other industries. The major participant firms are from financial, manufacturing, and extractive industries. The classification of sample firms by industries is presented in Table 2.

**Table 2: Classification of Sample by Industries**

| INDUSTRIES | FIRMS |
|---|---|
| Financial Industry (**Banking, Insurance, Support Services**) | 16 |
| Chemicals Industry (**Medicals & Pharmacy**) | 5 |
| Consumption Goods Industry (**Supermarkets, Foods, Beverages**) | 10 |
| Manufacturing Industry (**Equipment, Retailers, Engineering**) | 14 |
| Media Industry (**Publishing, Telecoms**) | 5 |
| Extractive Industry (**Oil, Gas, Energy, Mining**) | 13 |
| OTHERS INDUSTRIES | 17 |
| **TOTAL SAMPLE** | **80** |

The data availability for early years (GAAP years 2003-2004) had played an important role in determination of final sample size. Initially size of the sample was considered as whole FTSE 100 index firms. Notwithstanding to utilization of more than one database (Fame, Morningstar, Company Intelligence), data for the years of 2003-2004 were not available for all hundred firms listed in FTSE 100 index. Therefore, the data availability diminished the sample size from 100 to 80. The list of these 80 sampled firms is presented in the Appendix A in alphabetic order. Finally, a brief outline of final sample is provided in the Table 3.

**Table 3: A Brief Outline of Final Sample**

| Brief Outline of Final Sample | |
|---|---|
| Index | FTSE 100 index |
| Share of the Sample in the Index | **%90** |
| Number of the Sample Firms | 80 Firms |
| Total Market Capitalization of the Sample | £1.61bn |
| Total Assets of the Sample | £8,574bn |
| Total Turnover of the Sample | **£1,774bn** |
| Total Net Profit of the Sample | **£105.05bn** |
| Period of Observation | 2003-2006 years |
| Rows of data | 320 (4 years * 80 firms) |
| Selected Variables | • 10 Ratios (ROE, ROA, ROCE,PM, NAT, ST, CR, LQR, GR, BvMv), <br> • 5 Items (BV, MV, NI, TA, TL) |
| Data Source | Fama Database |

### 3.2. Data Description and Model

This study aims to capture IFRS adoption affects with two different analyses: First is "Test-A" which assesses impact of IFRS on value relevance; and second is "Test-B" which assesses impact of IFRS on key financial indicators. More importantly, the figures of 2003-2004 years represent the UK GAAP characteristics, and the figures of 2005-2006 years represent the new standards feature IFRS.

For Test-A, only market value of equity (MV), book value of equity (BV), and book value of net income (NI) are utilized as data input. The following hypothesis is established to find out whether there is any impact of IFRS adoption on value relevance.

- **$H_{01}$:** There is no statistically significant impact of IFRS adoption on value relevance.

To test the above hypothesis, I calculated the book-to-market value ratios under UK GAAP and IFRS with following formulas.

$$BvMv_{GAAP} = \sum_{t=2003}^{2004} \sum_{k=1}^{80} [BV_{t,k}/MV_{t,k}] \qquad BvMv_{IFRS} = \sum_{t=2005}^{2006} \sum_{k=1}^{80} [BV_{t,k}/MV_{t,k}]$$

After calculations of book-to-market value ratios, I applied Shapiro-Wilk and Lilliefors normality tests (Appendix C) to check whether variables are normally distributed or not. The t-test is then applied to the variables which are found to be normal. For non-normal variables non-parametric tests (Wilcoxon/Mann-Whitney test, Kruskal-Wallis test, Van Der Waerden test) were employed (Appendix D) to observe post adoption impact on value relevance. Besides, for empirical analysis I used Ohlson's (1995) model -which was used in prior researches by Hung and Subramanyam (2004) and Tsalavoutas et al (2012)- to capture the impact of IFRS.

$$MV_{i,t} = b_0 + b_1 BV_{i,t} + b_2 NI_{i,t} + \varepsilon_{i,t}$$

Here, $MV_{it}$ (dependent) and $BV_{it}$ (independent) are the market and book value of total shares of $i$ firm at time $t$ respectively. Similarly, $NI_{it}$ (independent) is the net income of $i$ firm at time $t$, and $\varepsilon_{it}$ is other value relevant information of firm $i$ at time $t$. To capture transmission impact of value relevance, I included a dummy variable (MODE) to Ohlson's (1995) model that represents the UK GAAP when it is zero, and IFRS when it equals 1.

$$MV_{i,t} = b_0 + b_1 BV_{i,t} + b_2 NI_{i,t} + b_3 MODE_{i,t} + \varepsilon_{i,t}$$

For Test-B, 9 ratios are calculated from gathered data and classified into three groups such as profitability, efficiency-liquidity, and capital structure. The profitability of firms is represented by ROE, ROCE, ROA, and PM ratios, whereas the efficiency-liquidity is represented by NAT, ST, CR, and LQR. Besides, the capital structure group embodies only GR ratio, which characterizes financial leverage of the sampled firms. On the other hand, 5 additional items (MV, BV, NI, TA, TL) and BvMv ratio are also included to the Test-B analysis. The abbreviation of these selected variables is presented in the Appendix B. Finally, the following hypothesis is established to find out whether there is any difference between variables gathered from financial reports of different accounting standards: UK GAAP and IFRS.

- **$H_{02}$:** There is no statistically significant difference between key financial ratios of UK GAAP and IFRS.

To check whether the variables are normally distributed, I applied Shapiro-Wilk and Lilliefors' normality tests (Appendix C). Then I applied t-test for those are found normal, and non-parametric tests (Wilcoxon/Mann-Whitney test, Kruskal-Wallis test, Van Der Waerden test) for non-normal variables (Appendix D) to determine if there are any difference between pre- and post-adoption periods.

To make robust and identify how ratios are affected (positively or negatively) I assigned a Forward Stepwise regression analysis with a dummy variable MODE.

$$Ratio_{i,t} = c_0 + c_1 ITEM_{i,t} + c_4 MODE_{i,t} + \varepsilon_{i,t}$$

Here, the dependent variable $Ratio_{i,t}$ stands for 9 selected ratios of a firm $i$ at time $t$. Independent variables are $ITEM_{i,t}$ that stands for balance sheet items (BV, NI, TA, and TL) of a firm $i$ at time $t$, and $MODE_{i,t}$ that is a dummy variable which equals 1 during IFRS, and reverts back to zero during GAAP periods.

# 4. Analysis & Results

## *4.1.* **TEST-A (Value Relevance Analysis)**

The aftermath of applied normality tests shows that MV, BV, NI, and BvMv variables are appeared as non-normal distributed. Therefore, t-test could not be employed, so non-parametric tests such as Wilcoxon/Mann-Whitney, Kruskal-Wallis, and Van Der Waerden tests were utilized to observe whether there is a statistically significant difference between pre and post adoption variables.

**Table 4:Results of Non-parametric Tests**

| Variables | Wilcoxon/Mann-Whitney | | Kruskal-Wallis | | Van Der Waerden | |
|---|---|---|---|---|---|---|
| | *Statistics* | *Sign.* | *Statistics* | *Sign.* | *Statistics* | *Sign.* |
| MV | 2.6495 | 0.0081[***] | 7.0229 | 0.0080[***] | 7.7772 | 0.0053[***] |
| BV | 1.1546 | 0.2482 | 1.3346 | 0.2480 | 1.7252 | 0.1890 |
| NI | 2.2797 | 0.0226[**] | 5.1997 | 0.0226[**] | 5.1623 | 0.0231[**] |
| BvMv | 2.7111 | 0.0067[***] | 7.3533 | 0.0067[***] | 7.2086 | 0.0073[***] |

*\*Significant at 10% level, \*\*: Significant at 5% level, \*\*\*: Significant at 1% level.*

Table 4 demonstrates the results of non-parametric tests. According to the results, statistically significant differences between MV and BvMV variables are found at 1% level. Likewise, NI variable also appears to be statistically significant at 5% level. The focus of this analysis was BvMv as it represents the value relevance. According to aftermath, the $H_{01}$ hypothesis is rejected for BvMv, and it is concluded that the value relevance is affected by transmission to the new regime.

Moreover, the aftermath of Ohlson's (1995) model with dummy variable (MODE) makes robust above findings as it is seen in the Table 5. The intercept, BV, and NI are found statistically significant at 1% level. Likewise, the dummy variable MODE is found significant at 10% level. This result concretes that MV depends to the independent variables of BV, NI, and MODE. My main focus here was whether MV depends to the dummy variable MODE, which represents UK GAAP when it is zero, and IFRS when it is 1. As coefficient of MODE is positive, then it can be concluded that IFRS has positive affected or improved value relevance. The $R^2$ is found 74.14% that confirms that majority part of value relevance can be explained by independent variables of BV, NI, and MODE.

### Table 5: Results of Ohlson's Model with Dummy Variable

| Variable | Coefficient | t-Statistic | Prob. | Estimation Equation: (MV dependent) | |
|----------|-------------|-------------|-------|-------------------------------------|--|
| | | | | MV=C(1)+C(2)*BV+C(3)*NI+C(4)*MODE | |
| **C** | 2.7737 | 2.3912 | 0.0175[***] | | |
| **BV** | 1.0638 | 18.8303 | 0.0000[***] | R-squared | 0.7414 |
| **NI** | 2.1126 | 9.5156 | 0.0000[***] | F-statistic | 244.6085 |
| **MODE** | 2.3226 | 1.4919 | 0.1369[*] | Prob. | 0.0000 |

*Significant at 10% level, ** Significant at 5% level, *** Significant at 1% level.*

After finding that IFRS has affected the value relevance, I employed univariate analysis and compared the means of book-to-market value ratios under UK GAAP and IFRS. In Table 6 mean values imply that value relevance has decreased under IFRS from 0.52 to 0.37. But this eye-balling comparison does not perturb my previous findings that value relevance is positively affected. On the other hand, Table 6 demonstrates a decrease of volatility of BvMv from 0.47 to 0.26. Besides, the skewness and kurtosis of distributions also show some signs of improvements. Furthermore, I demonstrated descriptive statistics of NI to show volatility of profits is increased nearly 38%, whereas the skewness and kurtosis of its distributions are improved.

### Table 6: Descriptive Statistics of BvMv ratios under UK GAAP and IFRS

| Book-to-Market Value Ratio (BvMv) | | | NI | |
|-----------------------------------|------|------|----------|----------|
| Standard | GAAP | IFRS | GAAP | IFRS |
| Mean | 0.52 | 0.37 | 1359.48 | 1945.81 |
| St Error | 0.04 | 0.02 | 281.11 | 389.55 |
| Median | 0.39 | 0.32 | 269.20 | 374.50 |
| Mode | #N/A | #N/A | #N/A | 466 |
| St Deviation | 0.47 | 0.26 | 3205.20 | 4441.53 |
| Sample Variance | 0.22 | 0.07 | 10.27+e6 | 19.72+e6 |
| Kurtosis | 7.86 | 4.51 | 19.71 | 12.95 |
| Skewness | 2.47 | 1.83 | 3.86 | 2.65 |
| Range | 3.00 | 1.53 | 27621 | 40812 |
| Minimum | -0.06 | 0.01 | -5047 | -14853 |
| Maximum | 2.94 | 1.54 | 22574 | 25959 |
| Sum | 67.72 | 48.54 | 176731.90 | 252955.86 |
| Count | 130 | 130 | 130 | 130 |
| Confidence Level (95%) | 0.08 | 0.04 | 556.19 | 770.73 |

### *4.2.* **TEST-B (Financial Ratios Analysis)**

In the Test-B, 9 mentioned ratios (profitability, efficiency-liquidity, and capital structure) are classified as pre- and post-adoption periods. My main focus here is to assess whether IFRS transmission has affected key financial indicators of firms. Before testing the $H_{02}$ hypothesis, I assigned Shapiro-Wilk and Lilliefors normality tests (Appendix C) to check whether ratios are normally distributed or not. Afterwards, I assigned t-test for those are found normal, and non-parametric tests (Wilcoxon/Mann-Whitney test, Kruskal-Wallis test, Van Der Waerden test) for non-normal variables (Appendix D) to determine if there are any difference between ratios of pre and post adoption periods.

The aftermath of assigned normality tests shows that all ratios are appeared as non-normal distributed, except ROCE, which appeared normally distributed under Shapiro-Wilk test. However, Lilliefors test disagrees with Shapiro-Wilk test, and verifies that ROCE is not normally distributed. Therefore, I assigned non-parametric tests (Wilcoxon/Mann-Whitney, Kruskal-Wallis, and Van Der Waerden tests) for all ratios to observe whether there is a statistically significant difference between ratios of pre and post adoption periods.

**Table 7: Results of Non-parametric Tests**

| Variables | Wilcoxon/Mann-Whitney | | Kruskal-Wallis | | Van Der Waerden | |
|---|---|---|---|---|---|---|
| | *Statistics* | *Sign.* | *Statistics* | *Sign.* | *Statistics* | *Sign.* |
| ROE | 2.2688 | 0.0233[**] | 5.1510 | 0.0232[**] | 3.3982 | 0.0653[*] |
| ROCE | 1.3764 | 0.1687[*] | 1.8968 | 0.1684[*] | 1.3497 | 0.1653[*] |
| ROA | 1.4078 | 0.1592[*] | 1.9841 | 0.1590[*] | 1.9618 | 0.1613[*] |
| PM | 1.4721 | 0.1410[*] | 2.1695 | 0.1408[*] | 2.3132 | 0.1283[*] |
| NAT | 0.0643 | 0.9487 | 0.0042 | 0.9481 | 0.0069 | 0.9338 |
| ST | 0.3447 | 0.7303 | 0.1194 | 0.7297 | 0.0004 | 0.9843 |

| | | | | | | | |
|---|---|---|---|---|---|---|---|
| CR | 0.7703 | 0.4411 | 0.5946 | 0.4406 | 0.6645 | 0.4150 |
| LQR | 0.9773 | 0.3284 | 0.9567 | 0.3280 | 1.1701 | 0.2794 |
| GR | 1.6263 | 0.1039[*] | 2.6476 | 0.1037[*] | 2.9907 | 0.0837[*] |

*Significant at 10% level, **Significant at 5% level, ***Significant at 1% level.*

Table 7 demonstrates results of non-parametric tests. All tests confirm that there are statistically significant differences between ROE and GR ratios at 5% and 10% levels respectively. Likewise, quite weak difference also appears in PM, ROA, and ROCE ratios at 10% level. Therefore, the $H_{02}$ hypothesis is rejected for all profitability ratios (ROE, ROCE, ROA, PM) and gearing ratio (GR). So, it is concluded that the IFRS has affected profitability and capital structure ratios of the firms. On the other hand, efficiency-liquidity ratios are failed to show significant results, so the $H_{02}$ hypothesis is accepted for them. It indicates that the IFRS has not affected efficiency and liquidity ratios of the sampled firms.

In addition to Test-B, I employed descriptive statistics to the ratios found to be affected by IFRS adoption. Table 8 finds that means of all ratios are increased after IFRS transmission, except ROE, which is slightly decreased from 39.07 to 38.81 by eye-balling comparison. Likewise, while volatilities of ROE and GR are sufficiently decreased, volatilities of others are increased. In terms of skewness and kurtosis of their distributions, ROE and ROA are worsened, while others are improved.

**Table 8: Descriptive Statistics of Significant Ratios**

| UK GAAP | | | | | | IFRS | | | | |
|---|---|---|---|---|---|---|---|---|---|---|
| *Standard* | *ROE* | *ROCE* | *ROA* | *PM* | *GR* | *ROE* | *ROCE* | *ROA* | *PM* | *GR* |
| Mean | 39.07 | 12.81 | 8.93 | 12.81 | 125.19 | 38.81 | 14.89 | 10.65 | 15.53 | 137.01 |
| St Error | 8.09 | 0.95 | 0.62 | 0.92 | 15.34 | 1.83 | 1.20 | 0.81 | 1.27 | 13.44 |
| Median | 21.34 | 11.34 | 8.15 | 10.54 | 80.45 | 25.50 | 13.33 | 9.25 | 13.17 | 89.92 |
| Mode | #N/A | 11.00 | #N/A | #N/A | 3.00 | 23.72 | #N/A | #N/A | #N/A | #N/A |
| St Deviation | 92.22 | 10.89 | 7.03 | 10.44 | 174.93 | 56.48 | 13.64 | 9.19 | 14.52 | 153.25 |
| Kurtosis | 58.61 | 0.85 | 1.52 | 3.18 | 11.92 | 22.97 | 0.73 | -.44 | -.79 | 1.46 |
| Skewness | 7.28 | 0.85 | 0.91 | 0.90 | 3.32 | 2.35 | 0.38 | 1.34 | 0.35 | 2.71 |
| Range | 872.41 | 51.73 | 41.23 | 78.14 | 980.31 | 734.42 | 85.39 | 67.94 | 124.5 | 896.50 |
| Minimum | -9.25 | -7.35 | -4.33 | -17.22 | 1.50 | 22.02 | -22.84 | -11.7 | -50.61 | 3.50 |
| Maximum | 863.16 | 44.38 | 36.90 | 60.92 | 981.80 | 712.40 | 62.55 | 56.22 | 73.88 | 900.00 |

| Count | 130 | 130 | 130 | 130 | 130 | 30 | 130 | 130 | 130 | 130 |

Since eye-balling comparisons do not have statistical significance, I utilized Forward Stepwise regression analysis for affected profitability and gearing ratios. This builds statistical background for these ratios to determine whether they affected positively or negatively.

**Table 9: Output of Forward Stepwise Regression with Dummy Variable**

| Ratios | C | MODE | BV | NI | TA | TL | $R^2$ | Sign. |
|--------|---|------|----|----|----|----|-------|-------|
| ROE | 41.74[(***)] | - | -0.73[(**)] | 1.42 | - | - | 0.0159 | 0.1289 |
| ROCE | 13.63[(***)] | 2.12[(*)] | -0.14[(***)] | 1.09[(***)] | -0.04[(***)] | - | 0.2230 | 0.0000 |
| ROA | 9.43[(***)] | 1.69[(*)] | -0.11[(***)] | 0.74[(***)] | -0.02[(***)] | - | 0.1959 | 0.0000 |
| PM | 12.95[(***)] | 2.00[(*)] | -0.29[(***)] | 1.15[(***)] | - | 0.01[(*)] | 0.1634 | 0.0000 |
| GR | 141.5[(***)] | 14.44 | -1.90[(***)] | - | - | -0.11[(*)] | 0.0512 | 0.0037 |

*Significant at 10% level, ** Significant at 5% level, *** Significant at 1% level.

In the case of ROE dependent variable, the Forward Stepwise regression analysis has included only constant term, BV, and NI. However the main focus here was the dummy variable MODE, which is excluded by the analysis. Moreover, $R^2$ is insufficient, and the regression itself lacks the significance. On the other hand, in case of other profitability ratios both MODE and whole regression achieve significance. Also, Table 9 clearly demonstrates positive relationship between ROCE-MODE, ROA-MODE, and PM-MODE, as coefficients are positive (2.12, 1.69, and 2.00 respectively) and significant at 10% level. But in case of GR as dependent variable, the MODE fails to be significant.

## 5. Conclusion

This paper has two contributions to IFRS adoption literature. The first is measuring impact of IFRS adoption on value relevance with TEST-A analysis, which hypothesizes $H_{01}$. Initially, all variables are penetrated through normality tests, and they failed to be normally distributed. Therefore, non-parametric tests are assigned to observe whether items of GAAP and IFRS reporting standards are statistically different. The result of this analysis rejects the $H_{01}$ hypothesis for BvMv, and it is concluded that the value relevance is affected by transmission to new regime. Subsequently to find out how the BvMv is affected, the Ohlson's (1995) model is assigned. The aftermath finds positive coefficient for dummy variable MODE, which indicates positive impact of IFRS on value relevance.

The second contribution is assessing impact of IFRS adoption on key financial indicators with TEST-B analysis that subjects 9 financial ratios and 5 balance sheet items. This test hypothesizes $H_{02}$ to explore whether financial indicators are affected by IFRS adoption. Initially normality tests are employed to find out whether ratios are normally distributed or not. As all ratios failed in normality tests, I assigned non-parametric tests to observe if ratios under two reporting standards are statistically different. The result of this analysis verifies that all profitability ratios (ROE, ROCE, ROA, PM) and gearing ratio (GR) have statistically significant differences (ROE and GR at 5% significance level, others are at 10% significance level). Therefore $H_{02}$ hypothesis is rejected for all profitability ratios (ROE, ROCE, ROA, PM) and gearing ratio (GR). So, it is concluded that the IFRS has affected profitability and capital structure ratios of the firms. However, for all efficiency and liquidity ratios the $H_{02}$ hypothesis is accepted which means IFRS has not affected these ratios.

In addition, to determine how these ratios are affected, I assigned a Forward Stepwise regression analysis with dummy variable MODE which equals zero for GAAP, and 1 for IFRS. In case of ROE dependent variable, the Forward Stepwise regression analysis includes only constant term, BV, and NI. It fails to include the dummy variable MODE, which was main focus here. Likewise the significance of regression and $R^2$ are also inadequate. However, in case of other profitability ratios both MODE and whole regression achieve significance. The output documents positive relationships between ROCE-MODE, ROA-MODE, and PM-MODE, as coefficients are positive (2.12, 1.69, and 2.00 respectively) and significant at 10% level. On the other hand, MODE fails to be significant in case of GR as dependent variable.

Lastly, the findings of TEST-A are in line with results of Adzis (2012) in Asia, Klimczak (2011) in Poland, Beke (2011) in Hungary, Hung and Subramanyam (2004) in Germany and Jermakowicz (2004) in Belgium. On the other hand, the findings of TEST-B are line with results of Terzi et al (2013) in Turkey, Tanko (2012) in Nigeria, Outa (2011) in Kenya, Blanchette et al (2011) in Canada, Lantto and Sahlstrom (2009) in Finland, and Rahmanova (2009) in USA.

# APPENDICES

**APPENDIX-A:** Sampled Firms

| | COMPANY NAME | | COMPANY NAME |
|---|---|---|---|
| 1 | Aggreko PLC | 41 | Kingfisher PLC |
| 2 | Amec PLC | 42 | Land Securities Group |
| 3 | Anglo American PLC | 43 | Legal & General Group PLC |
| 4 | Antofagasta PLC | 44 | Lloyds Banking Group |
| 5 | Arm Holdings PLC | 45 | Marks And Spencer Group PLC |
| 6 | Associated British Foods PLC | 46 | Meggitt PLC |
| 7 | Astrazeneca PLC | 47 | National Grid PLC |
| 8 | Aviva PLC | 48 | Next PLC |
| 9 | Babcock International Group | 49 | Old Mutual PLC |
| 10 | BAE Systems PLC | 50 | Pearson PLC |
| 11 | Barclays | 51 | Persimmon PLC |
| 12 | BG Group PLC | 52 | Prudential PLC |
| 13 | BHP Billiton PLC | 53 | Reckitt Benckiser Group PLC |
| 14 | BP PLC | 54 | Reed Elsevier PLC |
| 15 | British American Tobacco PLC | 55 | Rexam PLC |
| 16 | British Land CO | 56 | RIO Tinto PLC |
| 17 | BT PLC | 57 | Rolls-Royce Holdings PLC |
| 18 | Bunzl PLC | 58 | Royal Bank of Scotland Group |
| 19 | Burberry Group PLC | 59 | Royal Dutch Shell PLC |
| 20 | Capita PLC | 60 | Sabmiller PLC |
| 21 | Carnival PLC | 61 | Sage Group PLC |
| 22 | Centrica PLC | 62 | Schroders PLC |
| 23 | Compass Group PLC | 63 | Serco Group PLC |
| 24 | CRH PLC | 64 | Severn Trent PLC |
| 25 | Croda International PLC | 65 | Smith & Nephew PLC |
| 26 | Diageo PLC | 66 | Smiths Group PLC |
| 27 | Easyjet PLC | 67 | SSE PLC |
| 28 | G4S PLC | 68 | Standard Chartered |
| 29 | GKN PLC | 69 | Standard Life PLC |
| 30 | Glaxosmithkline PLC | 70 | Tate & Lyle PLC |
| 31 | Hammerson PLC | 71 | Tesco PLC |
| 32 | Hargreaves Lansdown PLC | 72 | Travis Perkins |
| 33 | HSBC Holdings | 73 | Tullow Oil PLC |
| 34 | IMI PLC | 74 | Unilever PLC |
| 35 | Imperial Tobacco Group PLC | 75 | Vedanta Resources PLC |
| 36 | Intertek Group PLC | 76 | Vodafone Group PLC |
| 37 | ITV | 77 | Weir Group PLC |
| 38 | J Sainsbury PLC | 78 | Whitbread PLC |
| 39 | John Wood Group PLC | 79 | William Hill PLC |
| 40 | Johnson Matthey PLC | 80 | WM Morrison Supermarkets PLC |

**APPENDIX-B:** Abbreviation of Variables Those Subjected to This Study

|  | **ABBREVIATION** |  |
|---|---|---|
| **RATIOS** | ROE | Return on Equity |
|  | ROCE | Return on Capital Employed |
|  | ROA | Return on Assets |
|  | PM | Profit Margin |
|  | NAT | Net Asset Turnover |
|  | ST | Stock Turnover |
|  | CR | Current Ratio |
|  | LQR | Liquidity Ratio |
|  | GR | Gearing Ratio |
|  | BV/MV | Book-to-Market Value Ratio |
| **ITEMS** | MV | Market Value |
|  | BV | Book Value |
|  | NI | Net Income |
|  | TL | Total Liabilities |
|  | TA | Total Assets |

**APPENDIX-C:** Results of Normality Tests

|  | GAAP | Shapiro-Wilk | | Lilliefors | | IFRS | Shapiro-Wilk | | Lilliefors | |
|---|---|---|---|---|---|---|---|---|---|---|
|  |  | *Statistics* | *Sign.* | *Statistics* | *Sign.* |  | *Statistics* | *Sign.* | *Statistics* | *Sign.* |
| **RATIOS** | ROE | 0.3004 | 0.0000 | 0.3357 | 0.0000 | ROE | 0.3677 | 0.0000 | 0.2634 | 0.0000 |
|  | ROCE | 0.9465 | 0.0001 | 0.1181 | 0.0001 | ROCE | 0.9841 | 0.1343 | 0.0814 | 0.0344 |
|  | ROA | 0.9553 | 0.0003 | 0.1006 | 0.0026 | ROA | 0.9221 | 0.0000 | 0.0953 | 0.0057 |
|  | PM | 0.9366 | 0.0000 | 0.0993 | 0.0031 | PM | 0.9058 | 0.0000 | 0.1133 | 0.0003 |
|  | NAT | 0.6704 | 0.0000 | 0.1806 | 0.0000 | NAT | 0.7866 | 0.0000 | 0.1378 | 0.0000 |
|  | ST | 0.5179 | 0.0000 | 0.3039 | 0.0000 | ST | 0.6356 | 0.0000 | 0.2611 | 0.0000 |
|  | CR | 0.1607 | 0.0000 | 0.4140 | 0.0000 | CR | 0.8056 | 0.0000 | 0.1197 | 0.0001 |
|  | LQR | 0.1533 | 0.0000 | 0.4120 | 0.0000 | LQR | 0.7464 | 0.0000 | 0.1613 | 0.0000 |
|  | GR | 0.5839 | 0.0000 | 0.2594 | 0.0000 | GR | 0.6911 | 0.0000 | 0.2216 | 0.0000 |
|  | BV/MV | 0.8087 | 0.0000 | 0.1681 | 0.0000 | BV/MV | 0.8946 | 0.0000 | 0.1307 | 0.0000 |
| **ITEMS** | MV | 0.5385 | 0.0000 | 0.2909 | 0.0000 | MV | 0.6027 | 0.0000 | 0.2636 | 0.0000 |
|  | BV | 0.4039 | 0.0000 | 0.3367 | 0.0000 | BV | 0.4942 | 0.0000 | 0.3075 | 0.0000 |
|  | NI | 0.5043 | 0.0000 | 0.3082 | 0.0000 | NI | 0.5615 | 0.0000 | 0.2975 | 0.0000 |
|  | TL | 0.3572 | 0.0000 | 0.4047 | 0.0000 | TL | 0.3127 | 0.0000 | 0.4316 | 0.0000 |
|  | TA | 0.3990 | 0.0000 | 0.4118 | 0.0000 | TA | 0.3406 | 0.0000 | 0.4325 | 0.0000 |

**APPENDIX-D:** Results of Non-Parametric Tests

| | GAAP | Wilcoxon/Mann-Whitney | | Kruskal-Wallis | | Van der Waerden | |
|---|---|---|---|---|---|---|---|
| | | *Statistics* | *Sign.* | *Statistics* | *Sign.* | *Statistics* | *Sign.* |
| **RATIOS** | ROE | 2.2688 | 0.0233 | 5.1510 | 0.0232 | 3.3982 | 0.0653 |
| | ROCE | 1.3764 | 0.1687 | 1.8968 | 0.1684 | 1.3497 | 0.2453 |
| | ROA | 1.4078 | 0.1592 | 1.9841 | 0.1590 | 1.9618 | 0.1613 |
| | PM | 1.4721 | 0.1410 | 2.1695 | 0.1408 | 2.3132 | 0.1283 |
| | NAT | 0.0643 | 0.9487 | 0.0042 | 0.9481 | 0.0069 | 0.9338 |
| | ST | 0.3447 | 0.7303 | 0.1194 | 0.7297 | 0.0004 | 0.9843 |
| | CR | 0.7703 | 0.4411 | 0.5946 | 0.4406 | 0.6645 | 0.4150 |
| | LQR | 0.9773 | 0.3284 | 0.9567 | 0.3280 | 1.1701 | 0.2794 |
| | GR | 1.6263 | 0.1039 | 2.6476 | 0.1037 | 2.9907 | 0.0837 |
| | BV/MV | 2.7111 | 0.0067 | 7.3533 | 0.0067 | 7.2086 | 0.0073 |
| **ITEMS** | MV | 2.6495 | 0.0081 | 7.0229 | 0.0080 | 7.7772 | 0.0053 |
| | BV | 1.1546 | 0.2482 | 1.3346 | 0.2480 | 1.7252 | 0.1890 |
| | NI | 2.2797 | 0.0226 | 5.1997 | 0.0226 | 5.1623 | 0.0231 |
| | TL | 1.1884 | 0.2347 | 1.4142 | 0.2344 | 1.6171 | 0.2035 |
| | TA | 1.1661 | 0.2436 | 1.3618 | 0.2432 | 1.9968 | 0.1576 |